\documentclass[12pt]{article}
\usepackage{chetnew}
\usepackage{xparse}
\usepackage{amssymb, amsmath}
\usepackage{xcolor}
\usepackage{tikz}
\usepackage{cite}
\usepackage[linktocpage]{hyperref}
\usepackage{mathtools}

\DeclareFontShape{OT1}{cmr}{mx}{n}%
    {<->cmr10}{}

\renewcommand{\bar}{\overline}

\def\Nequals#1{$\mathcal{N}{=}\,#1$}
\def\rank#1{\text{rank}\,#1}

\allowdisplaybreaks

\preprint{
NITEP 233}

\title{$S^1$ reduction of 4D \Nequals4 Schur index and \\[5mm]
3D \Nequals8 mass-deformed partition function}

\author{Tomoki Nakanishi$^{1,2\diamondsuit}$ and Takahiro
Nishinaka$^{1,2,3\clubsuit}$}
\affiliation{\smallskip  $^1$Department of Physics, Graduate School of Science\\
Osaka Metropolitan University, Osaka 558-8585, Japan\\
\bigskip
$^2$Nambu Yoichiro Institute of Theoretical and Experimental Physics (NITEP)\\
Osaka Metropolitan University, Osaka 558-8585, Japan\\
\bigskip
$^3$Osaka Central Advanced Mathematical Institute (OCAMI)\\
Osaka Metropolitan University, Osaka 558-8585, Japan
\emails{$^{\diamondsuit}$nakanishiphys@gmail.com, $^{\clubsuit}$nishinaka@omu.ac.jp}
}

\abstract{We study the compactification of 4D \Nequals4 SYM on $S^1$ from the
viewpoint of the superconformal index.
In the cases that the gauge group of the 4D SYM is $U(N)$ and $Usp(2N)$, the
resulting 3D theory is
believed to be the ABJM theory with the Chern-Simons level $k=1$ and $k=2$,
respectively. This suggests that the small $S^1$ limit of the
superconformal index of these 4D $\mathcal{N}=4$ SYMs is identical to
the sphere partition function of the ABJM theories. Using a recently observed relation
between the 4D and 3D R-charges for theories with twelve or more 
supercharges,
we explicitly confirm this identity in the Schur limit of the 4D index. Our result provides
a direct quantitative check of the relation between 4D $\mathcal{N}=4$ SYMs and 3D
$\mathcal{N}=8$ ABJM theories.
}

\begin{document}

\maketitle
\toc

\section{Introduction}
\label{sec:intro}

When a 4D $\mathcal{N}=4$ super
Yang-Mills theory (SYM) is compactified on $S^1$, it is believed to flow down to a 3D
$\mathcal{N}=8$ superconformal field theory (SCFT) in the deep infrared. One
explanation for this is the following. Performing the dimensional
reduction of the
4D SYM, one obtains a 3D
$\mathcal{N}=8$ SYM, which is not conformal. In particular, the 3D Yang-Mills coupling
triggers an RG-flow, along which the
kinetic term of the Yang-Mills field disappears. At the IR fixed point of the
RG-flow, the theory is described by an $\mathcal{N}=8$ SCFT.
In the cases that the gauge group of the 4D theory is $U(N)$ and
$USp(2N)$, the IR fixed point is believed to be described by the
$U(N)_{k}\times U(N)_{-k}$ ABJM theory for $k=1$ and $k=2$,
respectively \cite{Aharony:2008ug}. Note that, while the ABJM theory generically
preserves only $\mathcal{N}=6$ supersymmetry, it enhances to $\mathcal{N}=8$
for these values of $k$.

In general, when a 4D $\mathcal{N}\geq 2$ SCFT
$\mathcal{T}_\text{4D}$ compactified on $S^1$ flows to a 3D $\mathcal{N}\geq 4$
theory $\mathcal{T}_\text{3D}$ in the infrared, the small $S^1$ limit of the superconformal index of
$\mathcal{T}_\text{4D}$ is identical, up to a prefactor, to the $S^3$ partition function of
$\mathcal{T}_\text{3D}$. This has been confirmed
for various genuine 4D $\mathcal{N}=2$ SCFT \cite{Dolan:2011rp,
Gadde:2011ia, Imamura:2011uw, Bullimore:2014nla, Buican:2015hsa}. Therefore, it is extremely
natural to compare the small $S^1$ limit of the superconformal index of
4D $\mathcal{N}=4$ SYMs with the $S^3$ partition function of the
$\mathcal{N}=8$ ABJM theories. 

There has been, however, no explicit
comparison of the small $S^1$ limit of the 4D $\mathcal{N}=4$ index and
the 3D $\mathcal{N}=8$ partition function in the literature.
 One reason for this is
that, unlike for generic $\mathcal{N}= 2$ SCFTs, the superconformal
index of 4D $\mathcal{N}=4$ SYMs gives rise to a divergent prefactor of
the form
\begin{align}
 \left(\frac{1}{\beta}\right)^{\text{rank}\,G}~,
\label{eq:prefactor1}
\end{align}
in the small $S^1$ limit, where $\beta$ is the radius of the $S^1$ and
$G$ is the gauge group of the SYM
\cite{ArabiArdehali:2015ybk}.\footnote{This divergent factor is absent for genuine
$\mathcal{N}=2$ SCFTs in four dimensions. For such theories, the
partition function has a divergent factor $\exp\left(\frac{8
\pi^2}{\beta}(a-c)\right)$ instead, where $a$ and $c$ are the conformal
anomalies.} On the 3D side, this corresponds to the situation that a
sub-space of the moduli space is not lifted even when the 3D theory is
put on $S^3$, and therefore implies that the corresponding $S^3$
partition function is divergent.
When comparing the 4D index with the 3D partition
function, this divergent factor needs to be carefully identified in the
localization formula for the 3D partition
function. 

Another reason for the absence of 3D/4D comparison for 4D
$\mathcal{N}=4$ SYMs is that, along the RG-flow from four to three dimensions,
a part of the flavor symmetry of the 3D $\mathcal{N}=8$ theory is
accidental at the IR fixed point and has no 4D counterpart. Indeed,
while the R-symmetry $\mathfrak{so}(8)_R$
of the 3D theory is of rank four, the $\mathfrak{su}(4)$ R-symmetry of
the 4D $\mathcal{N}=4$ SYM is of rank three. For the 3D/4D comparison,
one needs to identify this accidental 3D global (sub-)symmetry.

Recently, we have solved the above two problems in the cases that the
gauge group $G$ of the 4D
$\mathcal{N}=4$ SYM is $G=U(N)$ and $G=USp(2N)$. Indeed, we have
identified in \cite{Nakanishi:2022fvr} which sub-algebra of the global
symmetry of the ABJM theory is accidental when it is obtained as the IR
CFT of the RG-flow from the 4D $\mathcal{N}=4$ SYM.\footnote{The results
of \cite{Nakanishi:2022fvr} is applicable to more general theories; It applies to ABJM theories obtained by the $S^1$
compactification of 4D $\mathcal{N}\geq 3$ SCFTs.}
Since the symmetry is accidental,
the mass parameter associated with it must be turned off until the
theory reaches the IR
fixed point. This constraint on the mass parameter gives rise to a
divergent prefactor of the $S^3$ partition function, which is naturally
 identified with the factor
\eqref{eq:prefactor1}. Hence, the results of \cite{Nakanishi:2022fvr} simultaneously
solved the first and second problems discussed in the previous two
paragraphs, and now there is no large obstacles to the 3D/4D comparison
for 4D $\mathcal{N}=4$ SYMs.

Let us look a little more into the divergent prefactor of the $S^3$
partition function here.
According to \cite{ArabiArdehali:2015ybk},
this divergent factor must be of the form
\begin{align}
 \Lambda^{\text{rank}\,G}~,
\label{eq:prefactor2}
\end{align}
where $\Lambda$ is the cut-off of the vacuum expectation value
(VEV) of scalar fields that parameterize (a sub-space of) the moduli
space of vacua. This implies that
a $(\text{rank}\,G)$-dimensional subspace of the moduli space is
un-lifted even after the theory is put on $S^3$ \cite{ArabiArdehali:2015ybk}. When the 3D theory
arises as the $S^1$-compactification of a 4D theory, such a flat
direction has a periodicity proportional to $\beta$. This means that
the cut-off parameter $\Lambda$ is proportional to $1/\beta$, and therefore
\eqref{eq:prefactor2} is identified as \eqref{eq:prefactor1}, up to a
numerical factor.

Given the above progress and observation, in this paper we explicitly compare the small $S^1$ limit of the
superconformal index $I$ of 4D $\mathcal{N}=4$ SYMs of gauge group $G$ with the $S^3$ partition
function $Z_{S^3}$ of the corresponding 3D $\mathcal{N}=8$ ABJM theories, focusing on the
cases of $G=U(N)$ and $G=USp(2N)$. 
For technical reasons, we only consider
the Schur limit of the 4D superconformal index, i.e., the Schur index $I_\text{Schur}$. As reviewed already, the
small $S^1$ limit of the 4D index gives rise to a power-law divergence \eqref{eq:prefactor1},
which is identified as a divergent prefactor \eqref{eq:prefactor2} of the $S^3$ partition
function. It is therefore natural to expect that an identity of the form
\begin{align}
\lim_{\beta \to 0} \beta^{\,\text{rank}\,G}\,I_{\text{Schur}} = \lim_{\Lambda \to \infty} \Lambda^{-\text{rank}\,G}Z_{S^3}
\end{align}
holds, up to a numerical prefactor. We explicitly show that this is
indeed the case for 4D $\mathcal{N}=4$ SYMs of gauge
group $G=U(N),\,USp(2N)$. Since our expression for the
small $S^1$ limit of the 4D Schur index is valid for any gauge group
$G$, one can generalize our discussions to the cases of $G\neq U(N),\,
USp(2N)$. We leave it for future work.

The rest of this paper is organized as follows. In section
\ref{sec:4dindex}, we briefly review the Schur limit of the
superconformal index of 4D $\mathcal{N}=4$ SYM for gauge group $G$,
including its useful expression recently discussed in
\cite{Hatsuda:2022xdv}. In section \ref{sec:s1reduction}, we consider the
small $S^1$ limit of the Schur index reviewed in section
\ref{sec:4dindex}. In section \ref{sec:comparison}, we compare the small $S^1$
limit of the 4D index with the $S^3$ partition function of the 3D
$\mathcal{N}=8$ ABJM theories. In section \ref{sec:conc}, we conclude
and discuss future directions.





\section{Schur index of 4D $\mathcal{N}=4$ SYMs}
\label{sec:4dindex}

Here, we briefly review the Schur limit of the superconformal index 
of 4D \Nequals4 
super Yang-Mills theories, which we will call the 4D \Nequals4 Schur
index below.

The superconformal index 
of a 4D $\mathcal{N}=4$ SCFT is generally defined by
\cite{Romelsberger:2005eg, Kinney:2005ej}
\begin{align}
	I
(p,q,t,u)=\mathrm{Tr}(-1)^Fp^{j_2-j_1+}q^{j_2+j_1-r}t^{r+R}u^f~,
\label{eq:index-def}
\end{align}
where the trace is taken over the space of local operators, and $(j_1,j_2)$
are $\mathfrak{so}(4)$ spins. The three charges $R, r$ and $f$ are linear combinations
of the Cartan generators of the R-symmetry $\mathfrak{su}(4)_R$, and
written as
\begin{align}
	R=\frac{1}{2}(\mathcal{R}^1{}_1-\mathcal{R}^2{}_2)~,\qquad
	r=\mathcal{R}^1{}_1+\mathcal{R}^2{}_2~,\qquad
 f=\mathcal{R}^1{}_1+\mathcal{R}^2{}_2+2\mathcal{R}^3{}_3~,
\label{eq:R}
\end{align}
where $R^I{}_J$ are generators of $\mathfrak{su}(4)_R$ with the
convention described in \cite{Nakanishi:2022fvr}.\footnote{One can take
an $\mathcal{N}=2$ superconformal sub-algebra, whose R-symmetry is
$\mathfrak{su}(2)_R\times \mathfrak{u}(1)_r$. The commutant of this
$\mathcal{N}=2$ R-symmetry algebra in $\mathfrak{su}(4)_R$ is
$\mathfrak{su}(2)_f$, and regarded as a flavor symmetry from the
$\mathcal{N}=2$ viewpoint. The charges $R, r$ and $f$ defined in
\eqref{eq:R} are associated with Cartan generators of
$\mathfrak{su}(2)_R,\, \mathfrak{u}(1)_r$ and $\mathfrak{su}(2)_f$,
respectively.}
For the above trace to be convergent, we impose $|p|,|q|,|t|<1,\,
|pq/t|<1$ and $|u|=1$.
%

The Schur limit
of the superconformal index is the limit
$t\rightarrow q$ \cite{Gadde:2011ik, Gadde:2011uv}, in which
the index preserves twice the number of supercharges
and the $p$-dependence of the index drops out. As a result,
one can set $p$ to any value in the Schur limit. In this paper, we will
set $p=q$ in the Schur limit, and therefore our expression for the Schur
limit of the $\mathcal{N}=4$ index is 
\begin{align}
 I_{\text{Schur}}(q,u) \equiv I(p,q,t,u)\big|_{t\to q,\,p = q}~.
\end{align}

The Lagrangian description for the $\mathcal{N}=4$ SYM implies that, for 
the 4D $\mathcal{N}=4$ SYM with gauge
group $G$, the $\mathcal{N}=4$ Schur index is evaluated as
\begin{align}
 I_{\text{Schur}}(q,u) = \int d\mu_G(z)\; \text{P.E.}\left[\,
 \left(i_{\mathcal{N}=2}^{\text{vec}}(q) +
 i_{\mathcal{N}=2}^{\text{hyp}}(q,u)\right)\chi_{\text{adj}}^{G}(z)\right]~,
\label{eq:N=4_G}
\end{align}
where 
\begin{align}
 i^{\text{vec}}_{\mathcal{N}=2}(q) \equiv -\frac{2q}{1-q}~,\qquad 
 i^{\text{hyp}}_{\mathcal{N}=2}(q,u) \equiv \frac{q^{\frac{1}{2}}}{1-q}(u+u^{-1})
\end{align}
are respectively single-letter indices for an $\mathcal{N}=2$ vector and an
$\mathcal{N}=2$ hyper multiplets, the plethystic exponential is defined
by $\text{P.E.}[f(q,u_i,z_i)] \equiv \exp\left(\sum_{k=1}^\infty \frac{f(q^k,(u_i)^k,(z_i)^k)}{k}\right)$, and $d\mu_G(z)$ and
$\chi_{\text{adj}}^{G}(z)$ are respectively the Haar measure on $G$ and
the character of the adjoint representation of $G$. To be explicit, the
latter two are expressed as
\begin{align}
	\mathrm{d}\mu_G
\equiv \frac{1}{|W(G)|}\prod^{\rank G}_{i=1}\frac{\mathrm{d}z_i}{2\pi iz_i}\,\prod_{\alpha\in\Delta_G}(1-z^\alpha)~,\qquad
	\chi_\text{adj}^G
\equiv \rank G+\sum_{\alpha\in\Delta_G} z^\alpha~,
\end{align}
where $|W(G)|$ is the order of the Weyl group of $G$, and $\Delta_G$ is
the root system of the Lie algebra $\mathfrak{g}$ of associated with
$G$.
In addition, for a root $\alpha=\sum_{i=1}n_i\alpha_i\in \Delta_G$ with $\alpha_i$ being simple roots,
 the factor $z^\alpha$ is defined by
\begin{align}
 z^\alpha 
\equiv \prod_{i=1}^{\text{rank} G}(z_i)^{n_i}~.
\end{align}
Putting the above expressions together in \eqref{eq:N=4_G}, one finds
\begin{align}
	I_{\text{Schur}}(q,u)
	 & =\frac{1}{|W(G)|}
	\left(\frac{(q;q)_\infty^2}{(q^{\frac{1}{2}}u^{\pm1};q)_\infty}\right)^{\!\!\rank G}\!
	\left(\prod^{\rank G}_{i=1}\oint_{|z_i|=1}\frac{dz_i}{2\pi iz_i}\right)
	\prod_{\alpha\in\Delta_G}
	\frac{(z^\alpha;q)_\infty(z^\alpha q;q)_\infty}{(z^\alpha q^{\frac{1}{2}}u^{\pm1};q)_\infty}~,
	\label{eq:N4SchurIndex1}
\end{align}
where $(a;b)_n:=\prod_{m=0}^n(1-ab^m)$ is $q$-Pochhammer symbol, and 
we used a shorthand $(xu^{\pm};q)_\infty \equiv
(xu;q)_\infty(xu^{-1};q)_\infty$ for an arbitrary fugacity $x$. 

It will turn out to be useful for us below to rewrite the
expression \eqref{eq:N4SchurIndex1} for the Schur index
further.
Following \cite{Hatsuda:2022xdv}, let us introduce
\begin{align}
\eta:=uq^{-\frac{1}{2}}~, 
\label{eq:xi}
\end{align}
and rewrite \eqref{eq:N4SchurIndex1} as
\begin{align}
	I_{\text{Schur}}(q, q^{\frac{1}{2}}\xi)
	=\frac{\eta^{\frac{\dim G}{2}}}{|W(G)|}
	\left(\frac{(q;q)_\infty^3}{\theta(\eta;q)}\right)^{2\,\rank G-\dim G}\,
	\left(\prod^{\rank G}_{i=1}\oint_{|z_i|=1}\frac{dz_i}{2\pi iz_i}\right)
	\prod_{\alpha\in\Delta_G}F(\eta,z^\alpha\eta^{-1};q)~.
	\label{eq:N4SchurIndex3}
\end{align}
where
\begin{align}
 \theta(x;q)&:=-x^{-\frac{1}{2}}(q;q)_\infty(x;q)_\infty(x^{-1}q;q)_\infty~,\qquad
 F(x,y;q):=\frac{\theta(xy;q)\,(q;q)_\infty^3}{\theta(x;q)\,\theta(y;q)}~.
\end{align}

In the 
next section, we will 
discuss the $S^1$ reduction
of the above Schur index of 4D $\mathcal{N}=4$ SYM for a general gauge
group $G$.

\section{$S^1$ reduction of the 4D index}
\label{sec:s1reduction}

Since the Schur index is identical to the partition function
of the 4D theory on $S^1\times S^3$ up to a normalization factor, its
small $S^1$ limit is expected to reproduce the $S^3$ partition function
of the 3D reduction of the 4D theory. 

In this section, we study the small $S^1$ limit of the Schur index of
4D $\mathcal{N}=4$ SYM with a general gauge group $G$. Unlike generic
$\mathcal{N}=2$ SCFTs, such a limit will lead to a power-law divergence for
these theories
\cite{ArabiArdehali:2015ybk}. For $G=U(N)$ and $USp(2N)$,
we will compare in the next section this divergent behavior with the $S^3$ partition
function of the 3D $\mathcal{N}=8$ theory obtained by dimensionally
reducing the 4D theory.

To that end, we first note that 
the fugacities, $q$ and $u$, in the Schur index of 4D $\mathcal{N}=4$
SYM are related to the radius $\beta$ of $S^1$ and a mass
parameter $M$ associated with a global $\mathfrak{su}(2)$ symmetry of the 3D reduction as \cite{Dolan:2011rp,
Gadde:2011ia, Imamura:2011uw, Bullimore:2014nla, Buican:2015hsa}
\begin{align}
	q=e^{-\beta}~,\qquad u=e^{-i\beta M}~.
	\label{eq:parameter}
\end{align}
Note that this and \eqref{eq:xi} implies 
\begin{align}
 \eta = e^{-i\beta\left(M +\frac{i}{2}\right)}
\label{eq:parameter2}
\end{align}
Here, the 3D reduction of the 4D $\mathcal{N}=4$ SYM is a 3D
$\mathcal{N}=8$ SCFT, whose $\mathfrak{so}(8)_R$ R-symmetry contains the
global $\mathfrak{su}(2)$ symmetry associated with $M$. The mass
parameters associated with the commutant of $\mathfrak{su}(2)$ inside
$\mathfrak{so}(8)_R$ are set to special values since we have taken the
Schur limit in four dimensions. These special values of the mass
parameters will be clear in the next section.

Below, we will substitute the parameter identification
\eqref{eq:parameter} in the expression \eqref{eq:N4SchurIndex3} for the Schur index
of $\mathcal{N}=4$ SYM with gauge group $G$, and then take the small
$S^1$ limit $\beta \to 0$. To illustrate our idea, we first discuss
lower rank cases, $G=U(1)$ and $G=U(2)$, in Sec.~\ref{subsec:example},
and then describe the general situation in Sec.~\ref{subsec:general}.

\subsection{Some examples}
\label{subsec:example}

We start with the case of $G=U(1)$. The expression
\eqref{eq:N4SchurIndex3} for the Schur index 
then reduces to
\begin{align}
	I_{\text{Schur}}
	=\eta^{\frac{1}{2}}\frac{(q;q)_\infty^3}{\theta(\eta;q)}  = \frac{1}{1-\eta^{-1}}
	\prod_{n=1}^\infty\frac{(1-q^n)^2}{(1-\eta q^n)(1-\eta^{-1}q^n)}~.
\end{align}
By substituting \eqref{eq:parameter} and \eqref{eq:parameter2}, this
index is expressed as
\begin{align}
 I_{\text{Schur}} &= \frac{1}{1-e^{i\beta(M+\frac{i}{2})}}
	\prod_{n=1}^\infty\frac{(1-e^{-\beta n})^2}{(1-e^{-i\beta(M +
 i(\frac{1}{2}+n))})(1-e^{i\beta(M+i(\frac{1}{2}-n))})}~,
\end{align}
which reduces in the limit $\beta \to 0$ to
\begin{align}
-\frac{1}{i\beta(M+\frac{i}{2})}\prod_{n=1}^\infty
 \frac{1}{\frac{\left(M+\frac{i}{2}\right)^2}{n^2} + 1}
= -\frac{\pi}{i\beta \sinh
 \left(\pi M + \frac{i\pi}{2}\right)}
= \frac{\pi}{\beta\cos (\pi M)}~.
\end{align}
%
Thus, we see that the small $\beta$ limit of the Schur index for
$G=U(1)$ is
expressed as
\begin{align}
 I_{\text{Schur}} = \left(\frac{2\pi}{\beta}\right)\frac{1}{2\cos
 (\pi M)}~, 
\label{eq:3Dlimit-U1}
\end{align}
where the sub-leading corrections are of $\mathcal{O}(\beta^0)$.

Next, let us consider the case $G=U(2)$.
The 4D \Nequals4 Schur index is then
\begin{align}
I_{\text{Schur}}
	=\frac{\eta^2}{2!}
	\oint_{|z_i|=1}\frac{dz_1dz_2}{(2\pi i)^2z_1z_2}
	\prod_{s=\pm 1}F(\eta,(z_1/z_2)^{s}\,\eta^{-1};q)~.
\end{align}
When $|q|<|y|<1$ and $|x|>1$, the function $F(x,y;q)$ has an infinite
series expansion
$F(x,y;q)=\sum_{m\in\mathbb{Z}}\frac{y^{-m}}{1-x^{-1}q^m}$
\cite{Zagier, Hatsuda:2022xdv}.
From \eqref{eq:xi} and the fact that $|q|<1,\,|u|=1$ and $|z_1|=|z_2|=1$, we see that
$|q|<|(z_1/z_2)^{s}\,\eta^{-1}|<1$ and $|\eta|>1$. Therefore, one can
use this series expansion to find
\begin{align}
	\prod_{s=\pm 1}F(\eta,(z_1/z_2)^{s}\,\eta^{-1};q)
	=\sum_{m,n \in\mathbb{Z}}\frac{(z_1/z_2)^{n-m}\eta^{-m-n}}{(1-\xi^{-1}q^m)(1-\eta^{-1}q^n)}~.
\end{align}
In the small $\beta$ limit, this reduces to 
\begin{align}
 	\frac{1}{\beta^2}\sum_{m,n \in\mathbb{Z}}\frac{(z_1/z_2)^{n-m}}{(-iM + \frac{1}{2}+m)(-iM + \frac{1}{2}+n)}~.
\end{align}
Since performing the contour integrals for $z_1$ and $z_2$ replaces
$(z_1/z_2)^{n-m}$ with $\delta_{n,m}$, we find that the small $\beta$
limit of the Schur index for $G=U(2)$
is expressed as
\begin{align}
 I_{\text{Schur}} =
 \frac{1}{2!\beta^2}\sum_{m\in\mathbb{Z}}\left(\frac{1}{-iM+\frac{1}{2}+m}\right)^2
 = \frac{1}{2!}\left(\frac{2\pi}{\beta}\right)^2\left(\frac{1}{2\cosh(\pi M)}\right)^2~,
\label{eq:3Dlimit-U2}
\end{align}
where the sub-leading corrections are of $\mathcal{O}(\beta^{-1})$.

\subsection{General gauge groups}
\label{subsec:general}

As shown above, the 3D reduction of the Schur indices of 4D
$\mathcal{N}=4$ SYM for gauge groups
$G=U(1)$ and $U(2)$ are expressed as \eqref{eq:3Dlimit-U1} and
\eqref{eq:3Dlimit-U2}, respectively. Here, we generalize them to an
arbitrary gauge group $G$.

To that end, first note that the Schur index \eqref{eq:N4SchurIndex3} can generally be expressed as
\begin{align}
I_{\text{Schur}}
	=\frac{\eta^{\frac{\dim G}{2}}}{|W(G)|}
	\left(\frac{(q;q)_\infty^3}{\theta(\eta;q)}\right)^{2\,\rank G-\dim G}
	\left(\prod^{\rank G}_{i=1}\oint_{|z_i|=1}\frac{dz_i}{2\pi iz_i}\right)
	\prod_{\alpha\in\Delta_G}\left(\sum_{m_\alpha\in\mathbb{Z}}\frac{(z^\alpha \eta^{-1})^{-m_{\alpha}}}{1-\eta^{-1}q^{m_\alpha}}	\right)~.
\end{align}
Taking the small $S^1$ limit 
$\beta\rightarrow0$, 
we 
obtain
\begin{align}
	\left.I_{\text{Schur}}\right|_{\beta\rightarrow0
}
	 & =\frac{1}{|W(G)|}
	\left(\frac{1}{\beta}\right)^{\rank G}
	\left(\frac{\pi}{\cosh(\pi M)}\right)^{2\,\rank G-\dim G}
 \nonumber
\\
	 & \times\left(\prod^{\rank G}_{i=1}\oint_{|z_i|=1}\frac{dz_i}{2\pi iz_i}\right)
	\prod_{\alpha\in\Delta_G}
	\left(\sum_{m_\alpha\in\mathbb{Z}}
	\frac{(z^\alpha)^{-m_\alpha}}{iM+\frac{1}{2}+m_\alpha}\right)~,
	\label{eq:integral}
\end{align}
where the sub-leading corrections are of
$\mathcal{O}(\beta^{1-\text{rank}\, G})$.

We will now show that the product over $\alpha$ in
\eqref{eq:integral} can be replaced by
$(\pi/\cosh(\pi M))^{\text{dim}\,G - \text{rank}\, G}$. Indeed, in terms
of the space of positive roots $\Delta_G^+$, this product is rewritten as
\begin{align}
 	\prod_{\alpha\in\Delta_G}
	\left(\sum_{m_\alpha\in\mathbb{Z}}
	\frac{(z^\alpha)^{-m_\alpha}}{iM+\frac{1}{2}+m_\alpha}\right) =
 \prod_{\alpha\in \Delta_G^+}\left(\sum_{m_\alpha,m_{-\alpha}\in
 \mathbb{Z}} \frac{(z^\alpha)^{-m_\alpha}
 (z^{-\alpha})^{-m_{-\alpha}}}{(iM+\frac{1}{2}+m_\alpha)(iM+\frac{1}{2}+m_{-\alpha})}\right)~.
\label{eq:product}
\end{align}
For $\alpha\in \Delta_{G}^+$, we change the variable from
$m_{-\alpha}$ to 
\begin{align}
 n_\alpha \equiv m_{-\alpha}-m_{\alpha} \qquad (\alpha\in\Delta_G^+)~,
\end{align}
and treat $m_\alpha$ and $n_\alpha$ as independent variables, so that
\eqref{eq:product} is expressed as
%
%
\begin{align}
	\prod_{\alpha\in\Delta_{G+}}\left(\sum_{m_{\alpha},n_\alpha\in\mathbb{Z}}
	\frac{(z^\alpha)^{n_\alpha}}{(iM+\frac{1}{2}+m_{\alpha})(iM+\frac{1}{2}+m_{\alpha}+n_\alpha)}\right)~.
	\label{eq:generalIntegral}
\end{align}
Note that, for any non-vanishing $n_\alpha$, the sum
\begin{align}
	\sum_{m_\alpha\in\mathbb{Z}}\frac{1}{(iM+\frac{1}{2}+m_\alpha)}\frac{1}{(iM+\frac{1}{2}+m_\alpha+n_\alpha)}
\end{align}
is identical to
\begin{align}
 \frac{1}{n_\alpha}\sum_{m_\alpha\in\mathbb{Z}}\left(\frac{1}{iM+\frac{1}{2}+m_\alpha}-\frac{1}{iM+\frac{1}{2}+m_\alpha+n_\alpha}\right)
 =0~.
\end{align}
This implies that \eqref{eq:generalIntegral} and therefore
\eqref{eq:product} can be rewritten as
\begin{align}
 	\prod_{\alpha\in\Delta_{G+}}\left(\sum_{m_{\alpha}\in\mathbb{Z}}
	\frac{1}{(iM+\frac{1}{2}+m_{\alpha})^2}\right) =
 \left(\frac{\pi}{\cosh(\pi M)}\right)^{\text{dim}\,G - \text{rank}\,G}~.
\end{align}
It is then straightforward to see that \eqref{eq:integral} is identical
to
\begin{align}
\left. I_{\text{Schur}}\right|_{\beta \to 0} =
 \frac{1}{|W(G)|}\left(\frac{2\pi}{\beta}\right)^{\text{rank}\,G}\left(\frac{1}{2\cosh
 (\pi M)}\right)^{\text{rank}\,G}~,
\label{eq:3Dlimit-G}
\end{align}
where the sub-leading corrections are of
$\mathcal{O}(\beta^{\,1-\text{rank}\,G})$~. This is the generalization
of \eqref{eq:3Dlimit-U1} and \eqref{eq:3Dlimit-U2} to an arbitrary gauge
group $G$.

\section{Comparison to the $S^3$ partition function of the ABJM theories}
\label{sec:comparison}

4D $\mathcal{N}=4$ SYMs are believed to be reduced down to 3D
$\mathcal{N}=8$ SCFTs by the $S^1$-compactification. Therefore, the 3D
reduction of the Schur index discussed in the previous section is
expected to be identical to the $S^3$ partition function of a 3D
$\mathcal{N}=8$ SCFT. 

In this section, we consider
4D $\mathcal{N}=4$ SYMs whose gauge group $G$ is $G=U(N)$ or $G=Usp(2N)$. In
this case, the corresponding
3D $\mathcal{N}=8$ SCFT is believed to be the
3D $U(N)_k\times U(N)_{-k}$ ABJM theory with $k=1$ for $G=U(N)$ and
$k=2$ for $G=USp(2N)$.
Below, we compare the $S^3$ partition
function of these ABJM theories with the small $\beta$ limit
\eqref{eq:3Dlimit-G} of the Schur index of the 4D $\mathcal{N}=4$ SYM of
gauge group $G=U(N),\, USp(2N)$.

One subtlety here is that, in the RG-flow from the 4D $\mathcal{N}=4$ SYM to
the ABJM theory, there is an accidental global $\mathfrak{u}(1)$ symmetry
that only exists in the IR fixed point. Indeed, the $\mathfrak{so}(8)_R$
global symmetry of the $\mathcal{N}=8$ ABJM
theory is of rank four while the $\mathfrak{su}(4)$ global symmetry of
4D $\mathcal{N}=4$ SYM is of rank three. Due to this accidental
symmetry, when flowing from four dimensions, one linear combination of
the three ($\mathcal{N}=2$ preserving) mass parameters of the ABJM theory are turned off, which gives
rise to a divergence in the $S^3$ partition function
\cite{Nakanishi:2022fvr}. This is a special phenomena for $\mathcal{N}\geq 3$
supersymmetry (in four dimensions) which does not occur for generic 4D 
$\mathcal{N}=2$ SCFTs.

Such a divergence has already been noticed in
\cite{ArabiArdehali:2015ybk} in the study of the small $S^1$ limit of the superconformal index
of 4D $\mathcal{N}=4$ SYM. Indeed, in the small $S^1$ limit, the
$\mathcal{N}=4$ index has a divergent prefactor
\begin{align}
\left(\frac{1}{\beta}\right)^{\text{rank}\,G}~,
\label{eq:beta}
 \end{align}
where $\beta$ is the radius of $S^1$.
This power-law divergence is interpreted in \cite{ArabiArdehali:2015ybk} to reflect the fact that the moduli
space of the corresponding 3D $\mathcal{N}=8$ SCFT has $\text{rank}\,G$ flat
directions even on $S^3$. This means that the $S^3$ partition function
of this 3D $\mathcal{N}=8$ SCFT is divergent; the path-integral of
scalar fields whose VEVs parameterize the flat directions give rise to a
divergence.
Specifically, when a cut-off $\sqrt{\Lambda}$ is introduced for the VEV of
each such (complex) scalar field, the partition function of
the 3D theory has a prefactor of the form
\begin{align}
 \Lambda^{\text{rank}\,G}~,
\label{eq:Lambda}
\end{align}
which is divergent in the limit $\Lambda\to\infty$.


The above observation given in \cite{ArabiArdehali:2015ybk} strongly suggests that the small $\beta$ limit of the Schur
index of a 4D $\mathcal{N}=4$ SYM of gauge group $G$ is identical to the
$S^3$ partition function of the corresponding 3D $\mathcal{N}=8$ SCFT,
under the identification of \eqref{eq:beta} with \eqref{eq:Lambda}. In
other words, we expect that the following relation holds up to a
numerical prefactor:
\begin{align}
 \lim_{\beta \to 0}\beta^{\text{rank}\,G}I_{\text{Schur}} =
 \lim_{\Lambda\to 0}\Lambda^{-\text{rank}\,G}Z_{S^3}~,
\end{align}
where $I_{\text{Schur}}$ is the Schur index of the $\mathcal{N}=4$ SYM,
and $Z_{S^3}$ is the $S^3$ partition function of the 3D $\mathcal{N}=8$
SCFT with the cut-off $\sqrt{\Lambda}$ introduced.
Below, we show this is indeed the case for $G = U(N)$ and
$G=USp(2N)$. Its generalization to more general gauge groups will be left
for future work.

\subsection{Partition function of mass-deformed ABJM theory}
\label{sec:massABJM}

We begin with the $S^3$ partition function of $U(N)_k \times U(N)_{-k}$
ABJM theory, which is evaluated by supersymmetric localization as
\cite{Kapustin:2009kz, Hama:2010av, Hama:2011ea}
\begin{align}
	&Z^{\text{ABJM}}_{S^3_b}(M,m)
	 =\int\frac{d^N\mu\ d^N\nu}{(N!)^2}e^{-i\pi k(\sum_i\mu_i^2-\sum_j\nu_j^2)}
	\prod_{i<j}(2\sinh(\pi (\mu_i-\mu_j))2\sinh(\pi (\nu_i-\nu_j)))^2 \nonumber \\
	 & \times\prod_{i,j}
	\left[
		s_{b=1}\left(\frac{i}{2}-\left(-\mu_i+\nu_j + \frac{-m_1-m_2+m_3}{2}\right)\right)\
		s_{b=1}\left(\frac{i}{2}-\left(-\mu_i+\nu_j+\frac{-m_1+m_2-m_3}{2}\right)\right)
	\right. \nonumber                                                                       \\
	 & \qquad\left.
		s_{b=1}\left(\frac{i}{2}-\left(\mu_i-\nu_j+\frac{m_1-m_2-m_3}{2}\right)\right)\
		s_{b=1}\left(\frac{i}{2}-\left(\mu_i-\nu_j+\frac{m_1+m_2+m_3}{2}\right)\right)
		\right]~,
\label{eq:S3}
\end{align}
where $s_{b=1}(z) \equiv  \prod_{n=1}^\infty
\left(\frac{n-iz}{n+iz}\right)^n$ is a specialization of the double-sine
function.
In \eqref{eq:S3}, the three parameters $m_1,m_2$ and $m_3$ are mass
deformation parameters that preserve a 3D $\mathcal{N}=2$
supersymmetry. Note that one needs to pick
an $\mathcal{N}=2$ sub-algebra of the $\mathcal{N}=8$ algebra to use the supersymmetric localization.
Following \cite{Chester:2021gdw}, we take $m_1,m_2$ and $m_3$ as follows. We start with the
manifest $\mathfrak{so}(6)_R\times \mathfrak{u}(1)_b$ symmetry, which is
believed to be enhanced to $\mathfrak{so}(8)_R$ symmetry in the
infrared. We choose an $\mathcal{N}=2$ supersymmetry whose R-symmetry is
$\mathfrak{so}(2)$. The commutant of the $\mathcal{N}=2$ R-symmetry in
the manifest global symmetry is $\mathfrak{so}(4) \times
\mathfrak{u}(1)_b\supset \mathfrak{so}(2)_1\times
\mathfrak{so}(2)_2\times \mathfrak{u}(1)_b$. We then take $m_1,m_2$ and
$m_3$ to be associated with $\mathfrak{so}(2)_1,\, \mathfrak{so}(2)_2$
and $\mathfrak{u}(1)_b$, respectively.

It was shown in \cite{Nakanishi:2022fvr} that, when the $S^3$ partition
function of the ABJM theory is
obtained as the small $\beta$ limit of the Schur index of the 4D $\mathcal{N}\geq 3$ SCFT, the three
mass parameters are constrained as\footnote{One can see this by setting
$\xi = -1$ and $b=1$ in Eq.~(3.8) of \cite{Nakanishi:2022fvr}. The fact
that $\xi=-1$ is needed
is proven in \cite{Nakanishi:2022fvr} while the condition $b=1$ corresponds to the Schur limit of the
4D index.}
\begin{align}
 m_1 = -\left(M+ \frac{i}{2}\right)~,\qquad m_2 = 0~,\qquad m_3 =
 \left(M-\frac{i}{2}\right)~,
\label{eq:Schur-mass-cond}
\end{align}
where $M$ is a mass deformation parameter corresponding to the fugacity
$u$ in the way as in \eqref{eq:parameter}. Note that this implies
\begin{align}
 m_1+ m_2+m_3 = -i~,
\label{eq:mass-constraint}
\end{align} 
which gives rise to a divergence in \eqref{eq:S3} since
$s_{b=1}(i-\mu_i-\nu_j)$ diverges when $\mu_i = \nu_j$. The
constraint \eqref{eq:mass-constraint} reflects the fact that the 3D
global $\mathfrak{u}(1)$ symmetry corresponding to the mass deformation
$(m_1+m_2+m_3)$ is accidental in three dimensions and has no 4D
counterpart.

Since \eqref{eq:Schur-mass-cond} leads to a divergent partition
function, let us introduce a small regularization parameter $\epsilon$ so that
\begin{align}
  m_1 = -\left(M+ \frac{i}{2}\right) + \epsilon~,\qquad m_2 = 0~,\qquad m_3 =
 \left(M-\frac{i}{2}\right) + \epsilon~,
\label{eq:Schur-mass-cond2}
\end{align}
which replaces the condition \eqref{eq:mass-constraint} with $m_1 + m_2
+ m_3 = -i + 2\epsilon$. Below, we will evaluate the leading
contribution to the partition function for $\epsilon \sim 0$.

Note that the regularization discussed above is
slightly different from the one we discussed around \eqref{eq:Lambda}. The reason for introducing a different regularization parameter
$\epsilon$ here is
that the localization formula \eqref{eq:S3} is available only when we take the
cut-off parameter $\sqrt{\Lambda}$ to be $\infty$. 
We will discuss later how the small mass parameter
$\epsilon$ is related to the cut-off parameter $\sqrt{\Lambda}$.

Plugging \eqref{eq:Schur-mass-cond} into \eqref{eq:S3} and using the
 identity $s_{b=1}\!\left(\frac{i}{2} -z\right) s_{b=1}\!\left(\frac{i}{2}+z\right) =
 \frac{1}{2\cosh (\pi z)}$, one obtains
\begin{align}
	Z_{S^3}(M)
	= i^{N^2}\int\frac{d^N\mu\,d^N\nu}{(N!)^2}e^{-i\pi k(\sum_i\mu^2-\sum_j\nu^2)}
	\frac{\prod_{1\leq i<j\leq N}\,(2\sinh\pi(\mu_i-\mu_j))^2\,
		(2\sinh\pi(\nu_i-\nu_j))^2}{
		\prod_{i,j=1}^N\,2\sinh\pi(\mu_i-\nu_j+\epsilon)\;\;2\cosh\pi(\mu_i-\nu_j-M)}~.
\end{align}
Using Cauchy's determinant formula
\begin{align}
	\frac{\prod_{1\leq i<j\leq N}(x_i-x_j)(y_i-y_j)}{\prod_{i,j=1}^N(x_i+y_j)}
	=\det\left(\frac{1}{x_i+y_j}\right)
	=\sum_{\rho}(-1)^{|\rho|}\prod_{i=1}^N\frac{1}{x_i+y_{\rho(i)}}
\end{align}
for 
$(x_i,y_j)=(e^{2\pi\mu_i},e^{2\pi(\nu_j-\frac{i}{2}-\epsilon)})$ and
$(e^{2\pi\mu_i},e^{2\pi(\nu_j+M)})$,
one obtains
\begin{align}
	Z_{S^3}(M)
	&= (-1)^{N^2}(-i)^N\int\frac{d^N\mu\, d^N\nu}{(N!)^2}e^{-i\pi
 k(\sum_i\mu^2-\sum_j\nu^2)}
\nonumber\\
&\qquad \sum_{\rho,\sigma}(-1)^{|\rho|+|\sigma|}\prod_{i=1}^N
\frac{1}{2\sinh\pi(\mu_i-\nu_{\rho(i)}+\epsilon)}
\frac{1}{2\cosh\pi(\mu_i-\nu_{\sigma(i)}-M)}~.
\end{align}
We further use the following formulae for Fourier transforms
\begin{align}
 \frac{1}{\sinh \pi x} = -i\int
 dp e^{2\pi ipx}\tanh \pi p~,\qquad \frac{1}{\cosh \pi x} =
 \int dq \frac{e^{2\pi i qx}}{\cosh \pi q}~,
\label{eq:Fourier}
\end{align}
to obtain
\begin{align}
	Z_{S^3}(M)
	&=
 \frac{1}{2^{2N}}\int\frac{d^N\mu\,d^N\nu\, d^Np\, d^Nq}{(N!)^2}\sum_{\rho,\sigma}(-1)^{|\rho|+|\sigma|}\prod_{i=1}^N
\frac{\tanh \pi p_i}{ \cosh \pi q_i}
\nonumber\\
&\qquad \times e^{\sum_{i=1}^{N}\big(-i\pi
 k\left(\mu_i^2-\nu_i^2\right) + 2\pi i
 p_i(\mu_i-\nu_{\rho(i)}+\epsilon) + 2\pi i q_i(\mu_i-\nu_{\sigma(i)}-M)\big)}~.
\end{align}
After performing the Gaussian integrals over $\mu_i$ and $\nu_i$, one obtains
\begin{align}
	Z_{S^3}(M)
	&= \frac{i^N}{(-4k)^N}\int\frac{d^Np\,d^Nq}{(N!)^2}\sum_{\rho,\sigma}(-1)^{|\rho|+|\sigma|}e^{2\pi i\sum_{i=1}^{N}\left( - q_iM +
 p_i\left(\frac{q_i-q_{\sigma^{-1}\circ \rho(i)}}{k}+ \epsilon\right)\right)}\prod_{i=1}^N
\frac{\tanh \pi p_i}{ \cosh \pi q_i}~.
\end{align}
Performing the $p_i$-integrals via \eqref{eq:Fourier}, one finds
\begin{align}
	Z_{S^3}(M)
	&=
 \frac{1}{(4k)^N}\int\frac{d^Nq}{(N!)^2}\sum_{\rho,\sigma}(-1)^{|\rho|+|\sigma|}e^{-2\pi
 i M \sum_{i=1}^Nq_i}\prod_{i=1}^N
\frac{1}{\sinh\pi\left(\frac{q_i-q_{\sigma^{-1}\circ\rho(i)}}{k} + \epsilon\right)}\frac{1}{\cosh \pi q_i}~.
\end{align}

Recall here that this partition function diverges in the limit
$\epsilon \to 0$, and we are interested in the leading divergent terms for
$\epsilon \sim 0$. Obviously, such leading terms arise only from the terms for
$\sigma^{-1}\circ\rho(i) = i$, or equivalently,
\begin{align}
 \sigma = \rho.
\end{align}
Thus, the leading divergent terms in the partition function is evaluated
as
\begin{align}
	Z_{S^3}(M)
	&\sim
 \frac{1}{(4k)^N}\left(\frac{1}{\sinh\pi\epsilon}\right)^N\int\frac{d^Nq}{(N!)^2}\sum_{\rho}e^{-2\pi
 i M \sum_{i=1}^Nq_i}\prod_{i=1}^N
\frac{1}{\cosh \pi q_i}~.
\end{align}
Using \eqref{eq:Fourier} and $\sinh\pi \epsilon \sim \pi \epsilon$, one
finally obtains
\begin{align}
	Z_{S^3}(M)
	&\sim
 \frac{1}{k^N N!}\left(\frac{1}{4\pi\epsilon}\right)^N\left(\frac{1}{\cosh\pi
 M}\right)^N~.
\label{eq:3D-result}
\end{align}
This is the leading divergent behavior of the $S^3$ partition function
of the ABJM theory in the small $\epsilon$ limit, under the constraints
\eqref{eq:Schur-mass-cond} on the mass parameters.

\subsection{Interpretation of the factor $1/\epsilon^N$}

The expression \eqref{eq:3D-result} implies that the ABJM theory on
$S^3$ has $N$ flat directions in the moduli space when the constraint
\eqref{eq:Schur-mass-cond} is imposed. This can be seen as follows.

Recall first that $\epsilon$ is a small mass parameter that we
introduced to regularize the divergence in $Z_{S^3}(M)$. Physically,
$\epsilon$ introduces a small real mass of zero modes of complex scalar
fields in (twisted) chiral
multiplets. Since the vacuum expectation values (VEV) of these chiral
multiplets parameterize a sub-space of the moduli
space of vacua, such a small mass parameter generally leads to a prefactor
$1/\epsilon^n$ in the partition function, where $n$ is the number of zero modes.

To see this explicitly, let us consider a complex scalar $\phi$ on $S^3$, whose
Lagrangian is given by \cite{Kapustin:2009kz, Hama:2010av, Hama:2011ea}
\begin{align}
 \mathcal{L} =\partial_\mu\bar{\phi}\partial^\mu \phi  +
 \epsilon^2\bar{\phi}\phi + 2i(\mathfrak{r}-1)\epsilon\bar{\phi}\phi + \mathfrak{r}(2-\mathfrak{r})\bar{\phi}\phi~,
\end{align}
where $\epsilon$ is a small real mass, and $\mathfrak{r}$ is the R-charge of the scalar field. The
$\mathfrak{r}$-dependent terms arise here since the background R-symmetry gauge field
is turned on to preserve supersymmetry on $S^3$.
 Now, let us focus on the zero mode $\phi = \phi_0$. Since the
 kinetic term $\partial_\mu\bar{\phi}\partial^\mu\phi$ vanishes for the
 zero mode, its path-integral is written as
\begin{align}
 \int d\bar{\phi}_0d\phi_0 \;e^{-\left(\epsilon^2 +
 2i(\mathfrak{r}-1)\epsilon + \mathfrak{r}(2-\mathfrak{r})\right)\int d^3x \bar{\phi}_0\phi_0}~.
\end{align}

Note that,  for the above path integral to be divergent, one needs to impose $\mathfrak{r}
= 0$ or $2$. In other words, only for these values of $\mathfrak{r}$,
the potential for the zero mode on $S^3$ is flat. The fact that we have
the divergent partition function \eqref{eq:3D-result} under the
condition \eqref{eq:Schur-mass-cond} implies that the mass of some chiral
multiplets with $\mathfrak{r}=0,2$ vanishes when
\eqref{eq:Schur-mass-cond} is imposed.
When $\mathfrak{r} =
0,2$, the above path-integral of the zero mode $\phi_0$ reduces to
\begin{align}
\int d\bar{\phi}_0 d\phi_0 \;e^{-\left(\epsilon^2 \mp
 2i\epsilon\right)V\bar{\phi}_0\phi_0} = \frac{\pi}{(\epsilon^2\mp 
 2i \epsilon)V}~,
\end{align}
where $V$ is the volume of $S^3$. Since we set the radius of
$S^3$ to one, $V=2\pi^2$. Substituting this and omitting the sub-leading
correction in the small $\epsilon$ limit, one can evaluate the above as 
\begin{align}
\int d\bar{\phi}_0 d\phi_0 \;e^{-\left(\epsilon^2 \mp
 2i\epsilon\right)V\bar{\phi}_0\phi_0} \sim \pm \frac{i}{4\pi \epsilon}~.
\label{eq:regularization1}
\end{align}

The expression \eqref{eq:regularization1} shows the leading divergence
of the path-integral of a
{\it single} (complex) zero mode $\phi_0$ corresponding to a flat direction on
the moduli space.
Therefore, when the moduli space has $n$ (complex) flat directions, the partition function with a regularization
parameter $\epsilon$ has a divergent factor of the form
\begin{align}
 \frac{1}{\epsilon^n}~.
\label{eq:factor1}
\end{align}
Comparing this with the divergent factor in \eqref{eq:3D-result}, we
now see that the moduli
space of $U(N)_k\times U(N)_{-k}$ ABJM theory  on $S^3$ has $N$ flat directions when the condition
\eqref{eq:Schur-mass-cond} on the mass parameters is satisfied.

\subsection{Comparison to the small $\beta$ limit of 4D index}

Let us finally compare \eqref{eq:3D-result} with the small $S^1$ limit
\eqref{eq:3Dlimit-G} of the 4D
index, along the lines that we mentioned at the beginning of this
section. For that, we first need to interpret the divergent factor
$1/\epsilon^N$ in \eqref{eq:3D-result} in terms of the cut-off
$\sqrt{\Lambda}$ for the VEV of scalar fields. 

Indeed, instead of introducing the small mass parameter $\epsilon$, one can regularize the
divergent path-integral of the zero modes by introducing the cut-off
$\sqrt{\Lambda}$ for the integration region. Such a regularization replaces
\eqref{eq:regularization1} with
\begin{align}
 \int_{-\frac{\sqrt{\Lambda}}{2}}^{\frac{\sqrt{\Lambda}}{2}} d\bar{\phi}_0 d\phi_0 = \Lambda~.
\end{align}
With this regularization, the existence of $N$ flat directions in the
moduli space leads to
the prefactor
\begin{align}
 \Lambda^N~,
\label{eq:factor2}
\end{align}
of the partition function instead of \eqref{eq:factor1}. This implies that, when introducing the cut-off $\Lambda$ instead of
the mass parameter $\epsilon$, the partition function
\eqref{eq:3D-result} is expressed as
\begin{align}
	Z_{S^3}(M)
	&\sim \frac{\Lambda^N}{k^N N!}\left(\frac{1}{\cosh\pi
 M}\right)^N~.
\label{eq:3D-result2} 
\end{align}
up to a numerical prefactor.

We now compare \eqref{eq:3D-result2} with the $S^1$-reduction
\eqref{eq:3Dlimit-G} of the Schur index of 4D $\mathcal{N}=4$ SYM of
gauge group $G$. As stated at
the beginning of this section, we focus on the cases $G=U(N)$ and
$G=USp(2N)$, in which cases the Chern-Simons level $k$ is set to $1$ and
$2$, respectively. In the case of $G=U(N)$, the 3D partition function is
evaluated as
\begin{align}
 	Z_{S^3}(M)
	&\sim \frac{\Lambda^N}{N!}\left(\frac{1}{\cosh\pi
 M}\right)^N = \frac{\Lambda^{\text{rank}\,U(N)}}{|W(U(N))|}\left(\frac{1}{\cosh\pi M}\right)^{\text{rank}\,U(N)}~.
\end{align}
When $\Lambda$ is identified with $1/\beta$, this is identical up to a
numerical factor
to the $S^1$ reduction \eqref{eq:3Dlimit-G} of the 4D index. In the case
of $G=USp(2N)$, the 3D partition function is given by
\begin{align}
 	Z_{S^3}(M)
	&\sim \frac{\Lambda^N}{2^N N!}\left(\frac{1}{\cosh\pi
 M}\right)^N = \frac{\Lambda^{\text{rank}\, USp(2N)}}{|W(USp(2N))|}\left(\frac{1}{\cosh\pi
 M}\right)^{\text{rank}\,USp(2N)}~.
\end{align}
Identifying $\Lambda$ with $1/\beta$, we see that the 3D
partition function is again identical to the 4D index up to a numerical
factor. Thus, we have seen that the relation
\begin{align}
\lim_{\beta \to 0} \beta^{\,\text{rank}\,G}\,I_{\text{Schur}} = \lim_{\Lambda \to \infty} \Lambda^{-\text{rank}\,G}Z_{S^3}
\end{align}
holds up to a numerical prefactor, at least for $G=U(N)$ and $G=USp(2N)$.




\section{Conclusions}
\label{sec:conc}

In this paper, we have studied the small $S^1$ limit of the
Schur limit of the superconformal index of 4D $\mathcal{N}=4$ SYMs of
gauge group $G=U(N)$ and $USp(2N)$, and compare the resulting
expression with the $S^3$ partition function of the 3D $\mathcal{N}=8$
ABJM theory of gauge group $U(N)_{k} \times U(N)_{-k}$ for $k=1$ and
$2$, respectively. On the one hand, our expression for the small $S^1$ limit of the 4D
Schur index is shown in Eq.~\eqref{eq:3Dlimit-G}, where $\beta$ is the
radius of $S^1$. On the other hand, that for the $S^3$ partition
function of the ABJM theory is shown in Eq.~\eqref{eq:3D-result} or
Eq.~\eqref{eq:3D-result2}, depending on whether we introduce the cut-off
$\sqrt{\Lambda}$ or the small mass $\epsilon$ as a regularization
parameter. Here, the reason for the $S^3$ partition function to be
divergent is that one linear combination of the mass parameters is
constrained as in Eq.~\eqref{eq:mass-constraint} when the 3D theory is
obtained as the $S^1$ compactification of a 4D theory. 
Under the identification of Eqs.~\eqref{eq:beta} and \eqref{eq:Lambda},
we have obtained a complete agreement between Eqs.~\eqref{eq:3Dlimit-G}
and \eqref{eq:3D-result2} up to a numerical prefactor.

One future direction would be to extend our result to more general pairs
of 4D and 3D theories with sixteen supercharges. For instance, there are
more 3D Chern-Simons matter theories that can have $\mathcal{N}=8$
supersymmetry \cite{Aharony:2008gk, Gang:2011xp}. Another future
direction would be to generalize our results on the Schur index to
the 4D superconformal index
with more general fugacities. For that, one needs to find a
generalization of the formula \eqref{eq:product} which has played a
crucial role in our discussions on the Schur index.

\ack{
	T.~Nakanishi’s research is partially supported by JST SPRING, Grant Number JPMJSP2139.
	T.~Nishinaka’s research is partially supported by JSPS KAKENHI Grant Numbers JP18K13547 and JP21H04993.
	This work was also partially supported by Osaka Central Advanced Mathematical Institute: MEXT Joint Usage/Research Center on Mathematics and Theoretical Physics JP- MXP0619217849.
}


%
%
%
%
%
%
%
%
%

\bibliography{reduction}

\end{document}